\documentclass[12pt,preprint]{aastex}

\begin{document}
\title{On the Difference in Statistical Behavior Between Astrometric and
Radial-Velocity Planet Detections}

\author{Andrew Gould}
\affil{Department of Astronomy, Ohio State University,
140 W.\ 18th Ave., Columbus, OH 43210, USA; 
gould@astronomy.ohio-state.edu}

\begin{abstract}
Astrometric and radial-velocity planet detections track
very similar motions, and one generally expects that the
statistical properties of the detections would also be similar
after they are scaled to the signal-to-noise ratio of the
underlying observations.  I show that this expectation is
realized for periods small compared to the duration of the
experiment $P/T\ll 1$, but not when $P/T\ga 1$.  At longer
periods, the fact that models of astrometric observations
must take account of an extra nuisance parameter causes
the mass error to begin deteriorating at $P/T\sim 0.8$, as compared
to $P/T\sim 1.0$ for RV.  Moreover, the deterioration is
much less graceful.  
This qualitative difference carries over
to the more complicated case in which the planet is being monitored
in the presence of a distant companion that generates an
approximately uniform acceleration.  The period errors begin
deteriorating somewhat earlier in all cases, but the situation
is qualitatively similar to that of the mass errors.
These results imply that to preserve astrometric discovery space at the
longest accessible orbits (which nominally have the lowest-mass
sensitivity) requires supplementary observations to identify or
rule out distant companions that could contribute quasi-uniform
acceleration.

\end{abstract}


\section{{Introduction}
\label{sec:intro}}

Astrometric and radial-velocity (RV) planet detections are, from
a mathematical standpoint, extremely similar.  In each, one
models 1-dimensional projections of Kepler orbits and attempts
to fit Kepler parameters.  In the limit of circular orbits, the planet
signature is encoded in the simple form
\begin{equation}
F(t;a_1,a_2,a_3) = a_1\sin(a_2 t + a_3),
\label{eqn:basicform}
\end{equation}
where 
\begin{equation}
a_1 = \alpha \quad {\rm (astrometry)},\qquad a_1 = K \quad {\rm (RV)},
\label{eqn:a1eq}
\end{equation}
are the astrometric and velocity semi-amplitudes for the two cases, 
and where
\begin{equation}
a_2 = {2\pi\over P}, \quad a_3 = \phi \qquad {\rm (astrometry + RV)}
\label{eqn:a23eq}
\end{equation}
designate the period $P$ and phase $\phi$, respectively.
One then goes on to combine information about the star's mass
and (for astrometry) its distance, to infer the planet mass $m$
(astrometry) or $m\sin i$ (RV), where $i$ is the inclination.
In the astrometric case, there are of course two such equations,
one for each direction in the plane of the sky, whose ratio
gives $\cos i$, and so permit one to break the $m\sin i$
degeneracy that plagues the intrinsically 1-dimensional RV measurement.

Because the form of equation (\ref{eqn:basicform}) is essentially
identical in the two cases, it is generally assumed that the
error properties, i.e., the relation between the measurement
errors and the derived-parameter errors, is also the same.
Of course, it is well known that the parameter errors have a different
dependence on semimajor axis $a$, system distance $D$, etc. Most
notably, with other parameters held fixed, astrometric sensitivity
increases linearly with $a$ whereas RV sensitivity declines as
$a^{-1/2}$.  But here I am referring to something else.  The differences
just mentioned all impact the final result because they change
the characteristic signal-to-noise ratio of the experiment,
\begin{equation}
{\rm SNR}\equiv {a_1\over\sigma}\sqrt{N},
\label{eqn:snrdef}
\end{equation}
where $N$ is the number of measurements and $\sigma$ is the
error in each measurement (assumed for simplicity to be
all the same).  Here I will show that the astrometric and RV
measurements have substantially different error properties
even when SNR is identical.

\section{{Minimum Variance Bound}
\label{sec:mvb}}

I will work within the framework of the minimum variance bound (MVB),
also frequently called the Fisher-matrix approximation.  As the same
approximation will be applied to both techniques, this will allow
me to highlight the difference between them.  Of course, if
equation (\ref{eqn:basicform}) really did fully represent both
techniques, there could not be any difference in their error
properties.  However, the true functional forms of the source motions are
actually
\begin{equation}
F(t;a_i) = a_1\sin(a_2 t + a_3) + \sum_{i=4}^n a_i t^{i-4}.
\label{eqn:trueform}
\end{equation}
For simplicity, I will continue to assume circular orbits, and
I will assume for the astrometric case that the orbit is seen
edge-on.  In fact, the latter is not much of an approximation
since in most cases the great majority of the information about
the mass and period comes from the major axis of the apparent
astrometric ellipse.  Finally, I will assume that the data are
taken uniformly in time, at intervals that are frequent compared
to the orbital period, and that they all have the same error, $\sigma$.

For a {\it single-planet}, i.e., a system composed of just a
star and a planet, in which the star does not suffer any other
accelerations, $n=4$ for the RV case, but $n=5$ for the astrometric
case.  For RV, $a_4$ is the systemic radial velocity of the system.
For astrometry, $a_5$ is the systemic proper motion of the system,
while $a_4$ is the zero point of that motion.

The introduction of the nuisances parameters $a_4$ or $(a_4,a_5)$ is
what distinguishes the two cases statistically.  The difference
appears only for $P\sim T$ or $P>T$, where $T$ is the duration of
the experiment.  I have implicitly assumed that $T,P\gg 1\,\rm yr$, and
this is the reason that I ignore parallax, which would be an
additional nuisance parameter for astrometry, but not for RV.

In the Fisher-matrix approximation, 
the inverse covariance matrix of the parameters,
$b_{ij}$, is given by
\begin{equation}
b_{ij} \equiv \sum_k {f_i(t_k)f_j(t_k)\over \sigma_k^2}
\rightarrow {N\over \sigma^2}\,{1\over T}\int_0^T d t f_i(t)f_j(t)
\label{eqn:bij}
\end{equation}
where
\begin{equation}
f_i(t)\equiv {\partial F(t)\over \partial a_i}.
\label{eqn:fi}
\end{equation}
One easily finds that for $P\ll T$, $a_1$ and $a_2$ are completely
uncorrelated from the other parameters, so that
\begin{equation}
{\sigma(a_1)\over a_1} \equiv {\sqrt{c_{11}}\over a_1} \rightarrow
 {1\over \sqrt{b_{11}}a_1} = \sqrt{2\over N}\,{\sigma\over a_1} =
{\sqrt{2}\over \rm SNR}.
\label{eqn:massfrac}
\end{equation}
That is, $\sigma(a_1)/a_1$, which is the fractional error in
astrometric or RV amplitude depends only on SNR, and in 
a very simple way.
Similarly,
\begin{equation}
{\sigma(P)\over P} = {\sigma(a_2)\over a_2} \rightarrow 
{1\over \sqrt{b_{22}}a_2} = {\sqrt{6}/\pi\over \rm SNR}\,{P\over T}.
\label{eqn:periodfrac}
\end{equation}

However, as $P$ approaches $T$, the parameters of primary interest
(mass and period) become correlated with the nuisance parameters.
To calculate this effect, I employ equations (\ref{eqn:bij}) and
(\ref{eqn:fi}), but using the full 7-parameter Kepler formalism,
not just the three Kepler parameters displayed in equations
(\ref{eqn:basicform}) and (\ref{eqn:trueform}).  That is, even
though the adopted orbits are circular and edge-on, I allow
for free fits to the eccentricity and inclination, and so for
correlations between these parameters (as well as the two remaining
Kepler parameters) and the parameters of interest (amplitude and period).
The effect of allowing for these covariances is to increase
the errors in semi-amplitude and period by modest amounts 
relative to what would be obtained using equation (\ref{eqn:trueform}).

\begin{figure}
\plotone{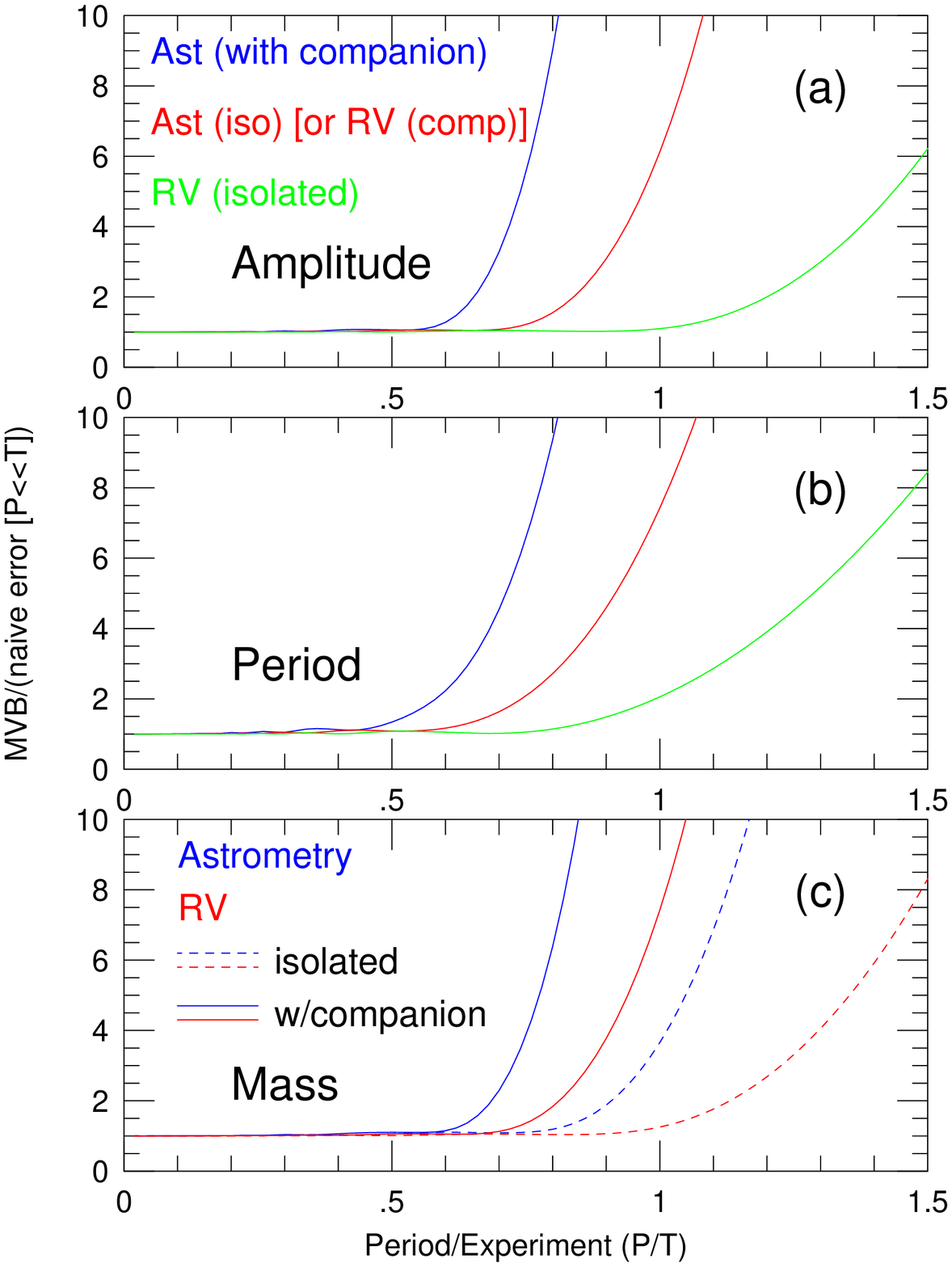}
\caption{\label{fig:apm}
Panels (a) and (b): 
ratios of true amplitude and period error to 
naive eqs.~(\ref{eqn:massfrac}) and (\ref{eqn:periodfrac})
as a function of $P/T$ and averaged over all phases $\phi$.  
Case of star with single planet is shown for RV
({\it green curve}) and astrometry ({\it red curve}).
They differ because astrometry has one
extra nuisance parameter (proper motion).  If the star suffers
an extra source of uniform acceleration from another
companion, there is an additional nuisance parameter for both
RV ({\it red curve}) and  astrometry ({\it blue curve}).  In all cases, RV
follows the naive equation to higher $P/T$ and degrades thereafter
more gracefully.  Panel (c): same
ratio for the mass error relative to eq.~(\ref{eqn:massfrac}).
Both the isolated ({\it dashed curves}) and
uniform-acceleration 
({\it solid curves}) cases are shown for RV
({\it blue curves}) and astrometry ({\it red curves}).
Note that the isolated-planet astrometry curve is no longer identical
to the uniformly-accelerated RV curve, as was true for amplitude
and period.
This is because the mass is derived from different combinations
of the amplitude and period for the astrometric and RV cases
(eqs.~[\ref{eqn:masspower}-\ref{eqn:masscovar}]).}
\end{figure}

Panels (a) and (b) of Figure \ref{fig:apm} show the ratios
of the true errors for the amplitude
and period, respectively, relative to the naive equations~(\ref{eqn:massfrac})
and (\ref{eqn:periodfrac}), for both the RV ({\it green})
and astrometric ({\it red}) cases.  In fact, as I will discuss in
\S~\ref{sec:phase}, for $P\ga T$, the errors depends on phase as
well as period.  Figure \ref{fig:apm} therefore shows the root-mean-square
of the errors, averaged over all phases.
Note that the RV amplitude errors 
follow the
naive form until $P/T\sim 1.1$ and then deteriorate relatively gracefully.
By contrast, the astrometric errors begin deviating at $P/T\sim 0.75$
and then deteriorate much more quickly.  For the period errors,
deterioration begins at $P/T\sim 0.85$ for RV and 
$P/T\sim 0.65$ for astrometry,
but the overall pattern is qualitatively similar.

The mass estimates for astrometry and RV depend on different 
combinations of amplitude and period,
\begin{equation}
m \propto a_1 a_2^n\qquad \biggl(n={2\over 3}\quad {\rm Ast};\quad
n=-{1\over 3}\quad {\rm RV}\biggr)
\label{eqn:masspower}
\end{equation}
Hence the fractional error in the mass (or $m\sin i$ in the case of RV)
is related to the errors in the fit parameters by
\begin{equation}
\biggl[{\sigma (m)\over m}\biggl]^2 = 
{c_{11}\over a_1^2} + 2n{c_{12}\over a_1 a_2} + n^2{c_{22}\over a_2^2}.
\label{eqn:masscovar}
\end{equation}
In the limit $P\ll T$, $\sigma(m)/m\rightarrow \sigma(a_1)/a_1$, but
for $P\ga T$, the period error and the correlations become important.
Figure \ref{fig:apm}c shows the results of calculations that
apply equation (\ref{eqn:masscovar}).

\section{{Additional Uniform Acceleration}
\label{sec:accel}}

Of course, the star may have more than one companion (planetary or 
otherwise), and one may imagine arbitrarily complicated configurations.
Here I restrict myself to the next level of complication, a second
companion that is sufficiently far away that its effect on the
star may be treated as uniform acceleration.  Even if no such
acceleration is identified, one might decide to fit for it on
the grounds that there {\it may} be such a companion that has
not been recognized.  I now ask how the inclusion of such an
additional nuisance parameter affects the precision of the 
physical parameters for the
planet that was previously treated as isolated.

The number of nuisance parameters is incremented by one in each case.
Referring to equation (\ref{eqn:trueform}), for RV, we now have
$n=5$, with $a_5$ being the systemic acceleration, while for
astrometry, $n=6$, with $a_6$ being the systemic proper-motion
acceleration.  
Figures \ref{fig:apm}a and \ref{fig:apm}b show the corresponding
ratios relative to the naive equations (\ref{eqn:massfrac})
and (\ref{eqn:periodfrac}).  In this case, RV is represented by
the {\it red curve}, while astrometry is represented by the
{\it blue curve}.  The fractional mass errors are shown by
{\it solid curves} in Figure \ref{fig:apm}c.

In this case, both deteriorate rapidly, but the mass-error deterioration
begins at $P/T\sim 0.65$ for astrometry and at $P/T\sim 0.75$
for RV.  One may anticipate that in other, yet more complicated
situations, RV will always perform like the naive equations
to higher $P/T$ than astrometry, simply because it has one
fewer nuisance parameter.

\section{{Phase Dependence}
\label{sec:phase}}

\begin{figure}
\plotone{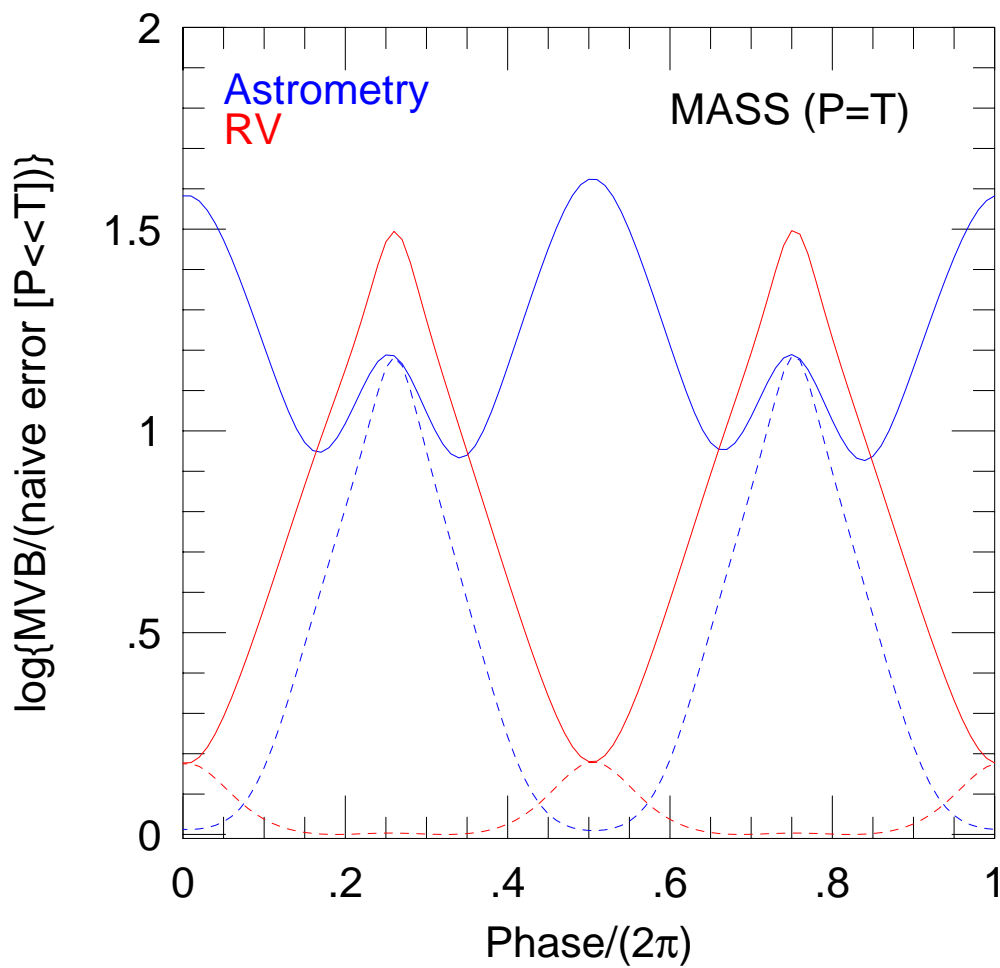}
\caption{\label{fig:phase}
Ratio of the true mass error to that given by the naive 
eq.~(\ref{eqn:massfrac}) as a function of phase $\phi$ for
the special case $P/T=1$.  When the ordinate in Fig.~\ref{fig:apm}
is large, the variations with phase are also large.  Hence,
Fig.~\ref{fig:apm} should be regarded as a guide to the statistical
properties of the experiments rather than a precise estimate of the
errors in any particular planet detection.}
\end{figure}

Figure~\ref{fig:apm} shows the root-mean-square errors averaged
over phase.  In fact, as $P/T$ grows and the rms errors deteriorate,
the variations with phase also increase.  Figure~\ref{fig:phase}
shows the phase variations of the mass errors for the particular
case that the period is exactly equal to the duration of the
experiment, $P/T=1$.  These large variations imply that
Figure~\ref{fig:apm} cannot be used to estimate the errors
in any particular case (except if the rms ratio is close to unity).
Rather, it should be used as a general guide to the reliability
of the experiment.  Any individual planet detection must be
analyzed based on the actual data and the parameters derived.

\section{{Discussion}
\label{sec:discussion}}

The results derived here are primarily of interest in regard
to future astrometric missions such as GAIA and SIM.  The
first point is that planets with periods that are even
slightly longer than the mission cannot be reliably detected
unless they are many times more massive than the nominal
thresholds of detection.  And second, if one is forced to allow for
uniform acceleration, then the same statement applies to 
$P/T\ga 0.7$.  Hence, to preserve the discovery space at the
longest accessible periods, it is critically important to
identify or rule out distant perturbers by supplementary data,
such as longer-term RV observations to look for large, distant
planetary or stellar companions, or perhaps AO observations
to find stellar companions.  This should be relatively
straightforward for SIM, with its limited number of high-precision
targets, but may be more difficult for GAIA, which is a survey
instrument.


\acknowledgments
I thank Wesley Traub and Scott Gaudi for valuable comments.
This work was supported by JPL contract 1239994 and NSF grant AST 042758.



\end{document}